\documentclass[reprint, superscriptaddress, amsmath,amssymb, aps, pra, prresearch, longbibliography, floatfix]{revtex4-2}

\usepackage{amsmath}
\usepackage{graphicx}
\usepackage{siunitx}
\usepackage{ulem}
\usepackage{todonotes}

\usepackage{bm}
\usepackage{dsfont}
\usepackage{tikz}
\usepackage{epsfig}
\usepackage{feynmf}
\usepackage{blindtext, rotating}
\usepackage{mathtools}
\usepackage{siunitx}
\usepackage{dsfont}
\usepackage{subcaption}
\usepackage{physics}
\usepackage{amsfonts}
\usepackage{xcolor}
\usepackage{ragged2e}
\usepackage{siunitx}
\usepackage{comment}
\usepackage{float} 
\usepackage{soul}
\usepackage{lipsum}
\usepackage{hyperref} 
\hypersetup{breaklinks=true, colorlinks=true, citecolor=blue, linkcolor=cyan, urlcolor=blue,filecolor=blue}

\usepackage{xcolor} 

\DeclareCaptionJustification{justified}{\justifying}

\DeclareSIUnit{\rad}{rad}

\begin{document}

\title{Feedback cooling of a levitated nanoparticle in a radially polarized vector beam trap}

\author{Felipe Almeida}
\email{felipe.almeida@ucl.ac.uk}

\author{M. Rademacher}
\affiliation{David Potter Institute for Quantum Information and Spacetime, University College London, Gower Street, WC1E 6BT London, UK}
\affiliation{Department of Physics and Astronomy, University College London, Gower Street, WC1E 6BT London, UK}
\author{J. M. H. Gosling}

\affiliation{Department of Physics and Astronomy, University College London, Gower Street, WC1E 6BT London, UK}
\author{P. F. Barker}
\email{p.barker@ucl.ac.uk}
\affiliation{David Potter Institute for Quantum Information and Spacetime, University College London, Gower Street, WC1E 6BT London, UK}
\affiliation{Department of Physics and Astronomy, University College London, Gower Street, WC1E 6BT London, UK}

\begin{abstract}
We demonstrate optical levitation and feedback cooling of a silica nanoparticle (radius = 78~nm) trapped within a radial vector beam (RVB) in high vacuum. Radial vector beams, when focused by a high-NA lens, produce a smaller focal spot than conventional Gaussian beams, yielding tighter optical potentials and potentially reduced photon-recoil heating rates. A vortex wave plate was used to generate the RVB using an aspheric lens (NA = 0.77) to form the trap. Cold-damping feedback was used to cool the center of mass motion in the radial and axial trap directions to temperatures in the mK range. At higher temperatures and pressures, additional trap frequencies arising from the nonlinear RVB optical potential are observed. This demonstration is a step toward exploiting structured light for levitated quantum optomechanics where these traps can be used to suppress bulk and motional heating.
\end{abstract}

\maketitle

\section{Introduction}

Levitated oscillators are promising platforms for exploring the limits of quantum mechanics and for probing the quantum nature of gravity~\citep{bose2017,Rademacher2019,Bose2025}. The manipulation of these levitated objects using optical fields has been successfully developed in the past few years, ultimately enabling cooling to the motional quantum ground state~\cite{Deli2020, Magrini2021, Piotrowski2023}. These optical traps are typically created by tightly focusing a single optical field with a Gaussian beam profile. For a particular trapped object, that is significantly smaller than the wavelength of light, the depth of the optical confining potential is determined by the light intensity at the focus, while the forces and subsequent trap frequencies are largely determined by the gradient of the intensity of the focused field.
Placing a levitated object into its ground state requires cooling of its motion, control of heating mechanisms, and maximization of the trap frequency.  Although trap frequencies can be increased with light intensity, an upper limit is imposed by the increased absorption in the levitated object, which although typically small, eventually leads to bulk heating and loss by melting with increased laser intensity \cite{Millen2014,Jnnemann2025}. Additionally, motional heating induced by the random scattering of photons from the levitating field, also dependent on light intensity, \cite{Seberson2020} eventually limits our ability to cool the motion of a levitated particle. This process also imposes significant constraints on the timescales over which quantum states can be created, manipulated, and witnessed within an optical trap~\citep{Romero-Isart2011}. Strategies to reduce bulk heating and recoil heating are, therefore, an active field of study and require the application of optical potentials beyond those produced by conventional Gaussian beams \cite{Almeida2023,Gajewski2025,Dago2024,Mlyn2026}. %comeback to it. Need proper referencing

The ability to independently modulate amplitude, phase and polarization of a light field  enables full customization of optical potentials. These so-called structured light modes \cite{He2022}  have been widely explored in the field of optical manipulation \cite{Dholakia2008}, optical and quantum communication \cite{Hu:20} and microscopy \cite{Xue2012}.  
Moreover, the combination of these higher order modes with spatially varying polarization distributions, known as vector beams \cite{Khonina2019,Jingwen2026}, has attracted interest in recent years. %https://journal.hep.com.cn/fop/EN/10.15302/frontphys.2026.042301
In particular, vector beams with a radially symmetric polarization profile have been of particular interest due to their distinctive properties when focused by a high numerical aperture lens. It has been shown that a radial vector beam can be focused to a focal spot considerably smaller than that of a Gaussian beam \cite{Dorn2003} which, in turn, yields steeper optical potentials and higher trap frequencies  for the same optical power when compared to conventional Gaussian beam traps \cite{https://doi.org/10.48550/arxiv.2510.05384, https://doi.org/10.48550/arxiv.2607.23833}. This makes these beams particularly attractive for optically levitated quantum-optomechanics experiments where both reduced bulk heating and recoil heating can be achieved \cite{https://doi.org/10.48550/arxiv.2510.05384}. Trapping in these beams has been verified experimentally in both liquids \cite{Huang2012,Donato2012,Kozawa2010} and air \cite{Michihata2009}.

In this work, we demonstrate optical levitation of a silica nanoparticle trapped in a radial vector beam at high vacuum. Using cold damping \cite{Gosling2026, Kremer2024}, we report on center of mass cooling to temperatures of $\SI{ 73 \pm 11 }{  \milli \kelvin}$ and $\SI{32 \pm 6}{  \milli \kelvin} $ along the $x$ and $z$ directions, respectively allowing us to levitate the nanoparticle in high vacuum (10$^{-6}$ mbar). We discuss the nonlinear effects arising from the resulting optical potential and outline the implementation of this type of trap for future levitated optomechanics experiments.
\section{Experimental setup}
\begin{figure}[h]
    \centering
    \includegraphics[width=1\linewidth]{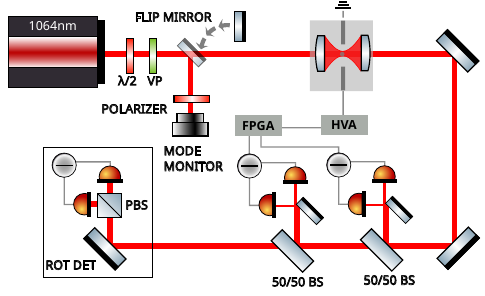}
    \caption{A schematic diagram of the experimental setup used to levitate nanoparticles in the radially polarized beam trap. The radial beam is created by a vortex plate (VP) and is characterized by a polarizer and a CCD (mode monitor) when the flip mirror is raised. After characterization, the radial beam is sent to an aspherical lens with high numerical aperture $(\mathrm{NA}=0.77)$ to form the optical trap. The scattered light from the nanoparticle is collected by a collector lens $(\mathrm{NA}=0.55)$. A balanced detection scheme is used to measure displacement of the nanoparticle and to cool the center of mass motion along the transverse $x$ and axial $z$ trap axes. A third channel (ROT DET) is also used to monitor the librational motion. }
    \label{fig:setup}
\end{figure}
The experimental setup used is shown in Fig. \ref{fig:setup}. The beam from a linearly polarized continuous wave (CW) laser (Coherent Mephisto), operating at a wavelength of $\lambda = \SI{1064}{\nano\meter}$, is collimated and converted into a radial vector beam (RVB) by passage through a vortex wave plate (VP in Fig. \ref{fig:setup}, THORLABS WPV10L-1064). The beam of approximately 100 mW power is then directed through an aspheric lens of numerical aperture $\text{NA}=0.77$ to create a tightly focused beam yielding a 3-D optical trap, also known as optical tweezers. The beam intensity and polarization profile are characterized using a CCD camera (mode monitor in the Fig. \ref{fig:setup}), which is accessed by raising the flip mirror. Silica nanoparticles (microParticles GmbH) of an approximate radius $R = \SI{78}{\nano \meter}$ are loaded into the vacuum chamber using nebulization \cite{Summers2008} and trapped at atmospheric pressure. Once trapped, the gas pressure is reduced initially to a pressure in the 10 mbar range for calibration of displacement measured by our detection system. An aspheric condenser lens, $\text{NA}=0.55$, %(Thorlabs A230TM-B) is then used to collect and collimate the scattered light 
is used to collect the light scattered by the nanoparticle in the forward direction. A $50/50$ beam splitter (BS) then divides the collected light into two branches, which are directed onto two orthogonally oriented D-mirrors that measure both the transverse and longitudinal displacement in the trap as a function of time. A third channel is also used to monitor the librational motion (labelled ROT DET in Fig. \ref{fig:setup}) of non-spherical particles which can also be trapped by the focused radial vector beam. 
%The observed mechanical frequencies are $\Omega_z = 2\pi \times \SI{38}{\kilo\hertz}$ and $\Omega_{x,y} = 2\pi \times \SI{104}{\kilo\hertz}$, where $z$ denotes the beam propagation direction, while $x$ and $y$ correspond to the transverse directions.
For feedback cooling of the translational motion, the two independent signals which are proportional to the displacement in the $x$ and $z$ axes are sent to a pair of field programmable gate arrays (FPGA, RedPitaya STEMlab 125-14) used to generate a feedback signal based on the particle's predicted velocity. Each feedback signal is obtained by applying a bandpass filter centered at the oscillator's frequencies with a phase delay that approximates the negative derivative of the displacement with time. %[] Goslin2026 + see cold damping papers 
These signals are then summed and amplified before being fed back to the two electrodes, positioned and configured, as in Ref.~\cite{Gosling2026}, inside the vacuum chamber. 
\section{Radially polarized vector beam}
\begin{figure}[h]
    \centering
    \includegraphics{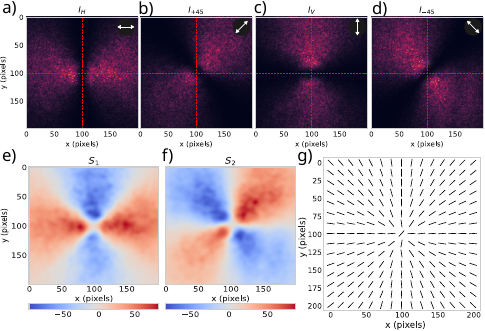}
    \caption{Polarization characterization of the radially polarized beam. The radial beam is filtered by rotating a polarizer from 0,45 , 90 and 135 degrees when positioned in front of the CCD camera. It generates four polarization states a) $I_{H}$, b) $I_{+45}$, c) $I_{V}$ and d) $I_{-45}$. The Stokes vectors e) $S_1=I_{H} - I_{V}$ and  f) $S_2=I_{+45} - I_{-45}$ are computed and the reconstructed polarization map of the radial beam is shown in g). }
    \label{fig:pol_reconstruction}
\end{figure}
Different techniques are used to generate radially polarized beams. These methods use intra-cavity modulation~\cite{Zhou2016}, % ref 50 of, 
diffractive and grating elements~\cite{Yang2024, Ye2023},  %diffractive elements, %gratings
or the superposition of Hermite Gauss (HG) beams. This includes $HG_{10} \bm{\hat{x}} + HG_{01} \bm{\hat{y}}$ or left- and right-circular helicoid beams defined by $LG_{01} ( i\bm{\hat{x}} + \bm{\hat{y}}) /\sqrt2 + LG_{0-1} ( i\bm{\hat{x}} - \bm{\hat{y}}) /\sqrt2$~\cite{Maurer2007}. %ref "tailoring of.."

In this work, we generate the radial vector beam with a q-plate \cite{Cardano2012} shown as VP in Fig. \ref{fig:setup}. This element can be viewed as a segmented half-wave plate with different optical axis directions. 
The polarization profile of the resulting beam is measured by inserting a polarizer in front of the CCD camera and rotating it to four configurations, with the corresponding transmission axes indicated in the top right of Fig. \ref{fig:pol_reconstruction}a-d and labeled $I_{H}$, $I_{+45}$, $I_{V}$,$I_{-45}$, respectively. These images are used to define the Stokes polarization parameters $S_1=I_{H} - I_{V}$ and $S_2=I_{+45} - I_{-45}$ as seen in Fig. \ref{fig:pol_reconstruction}e,f, respectively. These vectors can be used to reconstruct the orientation angle $\psi$ associated with the Poincar\'{e} sphere, such that $\psi = \frac{1}{2} \mathrm{arctan} (\frac{S_2}{S_1})$ %modulo pi
and the polarization profile of the radially polarized beam, as seen in Fig. \ref{fig:pol_reconstruction}.

\begin{figure}[h]
    \centering
    \includegraphics[width=1\linewidth]{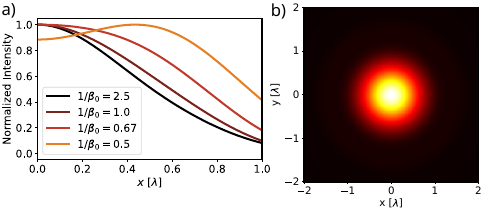}
    \caption{a) Transverse intensity for different filling factor $1/\beta_0$. Low filling factors prevent formation of a central intensity maximum. b) Transverse intensity profile for $1/\beta_0=2.5$ assumed in the simulations. }
    \label{fig:intensity_for_beta0}
\end{figure}

When tightly focused, the transverse component of the electric field of the radial beam is directed in the propagation direction. This creates a strong longitudinal component that is responsible for the bright spot in the focal region. In this plane, the electric field components can be described as~\cite{Youngworth2000, Kozawa2007}
\begin{subequations} \label{eq:E_field}
\begin{gather}
{E_\rho} = A \int_0^{\theta_{M}} \cos ^{1/2} (\theta) \sin (2\theta) P(\theta) J_1 (k\rho \sin \theta) e^{ikz \cos \theta} d\theta \\ 
{E_z}  = 2iA \int_0^{\theta_{M}} \cos ^{1/2} (\theta) \sin^2 (\theta) P(\theta) J_0 (k\rho \sin \theta) e^{ikz \cos \theta} d\theta
\end{gather}
\end{subequations}
\noindent where ${E_\rho}$ and ${E_z}$ are the electric field components in cylindrical coordinates centered around the focal plane, A is the normalized electric field, $\theta_{M} ={\arcsin(\text{NA/n})} $ is the maximum angle of the  optical pupil and $J_n$ is the first type of nth-order Bessel function. The apodization function, $P(\theta)$, is given by~\cite{Kozawa2007} %
\begin{equation}
    P(\theta) = \beta_0^2\frac{ \sin \theta}{\sin^2 \theta_{M}} \exp \left(  -\beta_0^2\frac{ \sin^2 \theta}{\sin^2 \theta_{M}} \right) L_0^1 \left( 2\beta_0^2\frac{ \sin^2 \theta}{\sin^2 \theta_{M}}\right)
\end{equation}
\noindent where $\beta_0=a/w_0$ denotes the inverse of the pupil filling factor, $a$ is the pupil radius and $w_0$ the incident beam waist in front of the lens. %Shall I plot the intensity fields for different trapping b_o? Discuss the b_0 importance in the arising of non-linearities?
Figure \ref{fig:intensity_for_beta0}a shows the transverse intensity in the $x$ direction for different values of the pupil filling factor, $1/\beta_0$.  %Figure 3 shows that the intensity mo sharper with increasing filling factor while at lower filling factors the absence of highly convergent rays prevents the formation of a central intensity maximum. % []. % [daniel]
For our experiment, we estimate a filling factor $1/\beta_0=2.5$, resulting in an intensity profile as shown in Fig. \ref{fig:intensity_for_beta0}b. 
\section{Results}
\begin{figure}[h]
    \centering
    \includegraphics[width=1\linewidth]{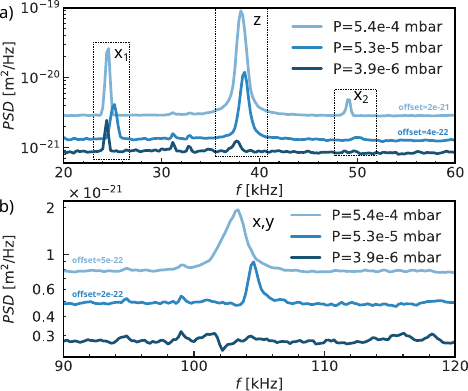}
    \caption{Recorded power spectral densities under feedback cooling for a levitated particle with nominal radius $R=\SI{78}{\nano\meter}$ at three different pressures. Motion is cooled to the noise floor level. The nonlinear peak denoted by $X_1$ dominates the spectra at $P=3.9 \times 10^{-6}$ mbar. }
    \label{fig:PSD}
\end{figure}

\begin{figure}[h]
    \centering
    \includegraphics[width=1\linewidth]{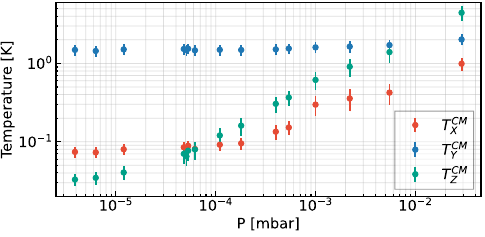}
    \caption{Center of mass temperature for different pressure. Minimum temperatures of $T^{CM}_{X}=\SI{ 73 \pm 11 }{  \milli \kelvin}$ and $T_{Z}^{CM} = \SI{32 \pm 6}{  \milli \kelvin} $ were found in the transverse and longitudinal directions, respectively.}
    \label{fig:temperature}
\end{figure}

Figure \ref{fig:PSD} shows the experimental displacement power spectrum density (PSD) when cold damping is applied along the $x$ and $z$ axes. Plots for three different pressures are shown. Unlike trapping in a linearly polarized Gaussian beam trap, which has three trap frequencies, the radially polarized beam trap is characterized by only two frequencies. One along the $z$ axis and another set of degenerate frequencies in the transverse directions along the $x$ and $y$ axes. The experimental power spectral density curves shown are produced by averaging $64$ time traces of $100$ ms each. They show the dominant mechanical frequencies at $\Omega_z = 2\pi \times \SI{38}{\kilo\hertz}$ and $\Omega_{x} = 2\pi \times \SI{104}{\kilo\hertz}$. 
Smaller sideband peaks are observed at $\SI{24.2}{\kilo\hertz}$ and $\SI{48.4}{\kilo\hertz}$, denoted by $\Omega_{x_1}$ and $\Omega_{x_2}$ in Fig.~\ref{fig:PSD}a. These are produced by the nonlinear trapping potential and are predicted by our simulation of the trapped motion in the transverse direction. This is discussed in Appendix B. While the main peaks $\Omega_{z}$ and $\Omega_{x,y}$ are suppressed to the noise floor level, the nonlinear peaks dominate the particle dynamics at $P=3.9 \times 10^{-6}$ mbar, which indicates that this transverse motion is not optimally cooled. Our simulations show that if both of these degrees of freedom are below $50~\text{K}$, these peaks would be strongly suppressed.
We verify the cooling efficiency of our scheme by computing the temperature of the main peaks' center of mass motion. This temperature, $T$, can be estimated for a trapped nanoparticle of mass $m$ through the equipartition theorem $m \langle \dot{q}^2\rangle = k_BT$, 
where $\langle \dot{q}^2\rangle$ is obtained by integrating the power spectral density, $\langle \dot{q}^2\rangle= \int_{-\infty}^{\infty} \Omega^2 \mathrm{S_{qq}}( \Omega) d \Omega /2\pi$ \cite{Hebestreit2018}. %ref Cavity Optomechanics, Aspelmeyer 
The motional temperatures can be seen in Fig. \ref{fig:temperature} as a function of gas pressure, $P$. The minimum temperatures achieved in the experiment were $T^{CM}_{X}=\SI{ 73 \pm 11 }{  \milli \kelvin}$ and $T_{Z}^{CM} = \SI{32 \pm 6}{  \milli \kelvin}$, as we reach the noise floor level of our detection scheme. The temperature of the $y$ mode has also been computed and although no active feedback has been performed, we measure temperatures as low as $\SI{ 1.4 \pm 0.2 }{\kelvin}$ by optimizing feedback cooling settings at $0.1 \text{ mbar}$.

\begin{figure}
    \centering
    \includegraphics[width=1\linewidth]{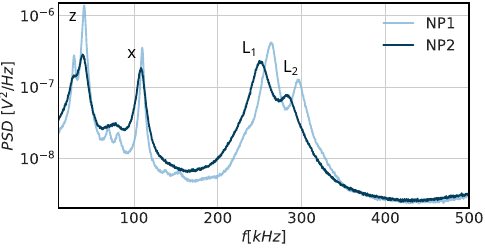}
    \caption{Power spectral density recorded on the rotation detector for two nanoparticles. The labels NP1 and NP2 denote two different particles trapped in the RVB at a pressure of $5 \text{ mbar}$ and $12 \text{ mbar}$, respectively. The peaks $L_1$ and $L_2$ are likely to be due to librational motion of non-spherical particles in the RVB. No feedback cooling was applied to these particles.}
    \label{fig:lib_detection}
\end{figure}

At the trap center, the highly focused radial vector beam is polarized along the propagation direction ($z$). The long axis of a non-spherical particle is therefore expected to align with the optical axis, and the trapped particle should undergo librational motion about it.  Fig.~\ref{fig:lib_detection} shows the power spectral density recorded on the rotation detector for two nanoparticles. NP1 and NP2 denote different particles trapped in the RVB at a pressure of $5 \text{ mbar}$ and $12 \text{ mbar}$, respectively. 
The peaks labeled in Fig.~\ref{fig:lib_detection} as $L_1$ and $L_2$ indicate the librational motion of non-spherical particles \cite{Zieliska2023}. Similar motion has been observed in traps formed by linearly polarized Gaussian beams, where, in contrast to the RVB, the polarization is predominantly orthogonal to the propagation direction \cite{Rashid2018}.
The non-spherical particles were also less stable than spherical ones. Without feedback, spherical particles remained trapped down to $10^{-2}~\text{mbar}$, while non-spherical particles left the trap at approximately $3 \times 10^{-1}~\text{mbar}$.

 %The highly focused radial vector beam is polarized along the propagation direction ($z$) at the trap center. It is expected that the long axis of a nonspherical particle would align along this optical axis leading to librational motion of the trapped particle. This type of motion is observed in linearly polarized Gaussian beam traps where the polarization is predominantly orthogonal to the direction of propagation \cite{Rashid2018}.  
%The setup also allows us to measure the constrained rotational/librational motion of non-spherical nanoparticles levitated in the trap.  As the polarization in the focal plane of the vortex beam is the along the axial trap direction, the long axis of an ellipsodial particle should align along this axis. At least two librational frequencies should be observed and these can be observed in Figure \ref{fig:lib_detection}. %which are plots of the PSD from the rotational detector. %F: Although I have plots refering the ROT detections, Fig. 6 is a plot of X-Channel detection
%The plot in red are for the PSD of a near spherical particle for comparison with the plots of particle 1 and 2 that show two librational peaks in the 250 to 300 kHz range. The PSD where taken at pressures of 10 mbar where trapping without feedback is stable. 

\section{Discussion and Conclusions}

We have demonstrated feedback cooling of levitated silica nanoparticles in high vacuum trapped by a radially polarized vector beam focused by a high numerical aperture lens. This work is a first step toward utilizing these beams in levitated optomechanics, where both bulk heating and recoil heating could be reduced by the larger intensity gradients in this type of trap. We have shown that the techniques typically used for Gaussian beam trapping and cooling can be used to detect and cool the motion of the particle in the trap. We demonstrate that center of mass temperatures in the tens of mK range can be achieved using the same detection and cooling strategies typically used for Gaussian beam experiments. This is possible even though collection of the scattered light is not optimized in this system \cite{https://doi.org/10.48550/arxiv.2607.23833}.  Significantly lower temperatures could be achieved by operating at lower pressures, by cooling all three degrees of freedom, and by matching the scattered light from the axially polarized light in the focal region to a split detection system. Within the uncertainties of our experiment, we find good agreement with the numerical estimates of the primary axial $z$ trap frequency and also the degenerate trap frequencies along the transverse $x$ and $y$ directions. The additional motional peaks observed are consistent with nonlinearities in the transverse optical potential. These were also observed in our simulations of the motion in the potential. While these additional peaks were reduced by cooling in the optical potential, as we only cool in the $z$ and $x$ directions, they are not completely suppressed. These nonlinearities, which are usually undesirable in levitated optomechanics experiments, could be significantly reduced by cooling all three motional degrees of freedom. On the other hand, non-Gaussian states of motion induced by non-harmonic potentials \cite{Filip2015} could be created by enhancing these effects using radially polarized vector beams. Here, potentials with even larger nonlinearities could be created by changing the filling factor as indicated by focused intensity profiles shown in Fig.~\ref{fig:intensity_for_beta0} which are proportional to the optical potential.  Lastly, we also measured additional higher-frequency peaks which we attribute to induced librational motion of non-spherical particles. While we have focused on trapping in the Rayleigh regime, trapping of larger particles in the Mie regime, where lower recoil heating is predicted \cite{https://doi.org/10.48550/arxiv.2510.05384}, is also feasible and deserves further experimental study. 
\section*{Acknowledgments}
F.A., M.R., J.M.H.G. and P.F.B. acknowledge funding from the
EPSRC via Grant Nos. EP/S000267/1 and EP/W029626/1.
M. R. acknowledges support by the Wellcome Trust Early-Career Award 331016/Z/25/Z. We acknowledge useful discussions with Antonio Pontin.
\section*{Appendix A: Detection Calibration }
The calibration of the detection system relies on an assumed linear dependency between the optical signal and the particle displacement. Following from the equation of motion for a trapped nanoparticle in the $q$-direction, the power spectral density is given by:
\begin{equation} \label{eq:psd}
    S_{qq} = \frac{2\gamma k_BT}{m\big[ (\Omega^2-\Omega_q^2)^2 + \gamma^2\Omega^2 \big]}
\end{equation}
\noindent where $\gamma$ is the oscillator damping rate and $T=293 K$ the environmental temperature.
As detection follows from an electrical signal, a calibration factor is necessary such that $S_{V_qV_q} (\Omega) = c^2_{q} S_{qq} (\Omega)$ where $S_{V_qV_q} (\Omega)$ represents the actual quantity measured in the laboratory and $c^2_{q}$ represents the calibration factor converting from $\mathrm{V^2/Hz}$ to $\mathrm{m^2/Hz}$. 
%Following Ref. [], %Hebestreit et al 2018
We record the $64$ time traces of $100 \text{ms}$ each and take an average PSD that is used to fit Eq. \ref{eq:psd}, which outputs $c^2_{\text{q}}$, $\Omega_i$ and $\gamma$. The coefficients were found to be: $C_x = (1.6 \pm 0.3) \times 10^6 $ V/m, $C_y = ( 5.5\pm0.8) \times 10^5$ V/m and $C_z = (1.4 \pm 0.2) \times 10^6$ V/m. 

To check the validity of these numbers, we model the trapped nanoparticle as a sphere, which is supported by the almost equal ratio of damping rates in the $x$ and $z$ direction given by  $\gamma_x/\gamma_z =1.1$. Additionally, there are no visible librational modes in the PSD \cite{Rademacher2022, Rademacher2026}. We model the gas damping associated with the sphere to be $\gamma=\frac{m_{\mathrm{gas}} v_{\mathrm{th}} P}{k_BT_0 \rho R}(1+\pi/{8})$ \cite{Cavalleri2010, Pontin2020}. %Cavalieri and Pontin
The fitted values allow us to extract an experimental measured radius of $R_{\text{exp}}= (76 \pm 8) \text{ nm}$, which is in agreement with the nominal size provided by the manufacturer.

\section*{Appendix B: Numerical Simulations}

\begin{figure}
    \centering
    \includegraphics[width=1\linewidth]{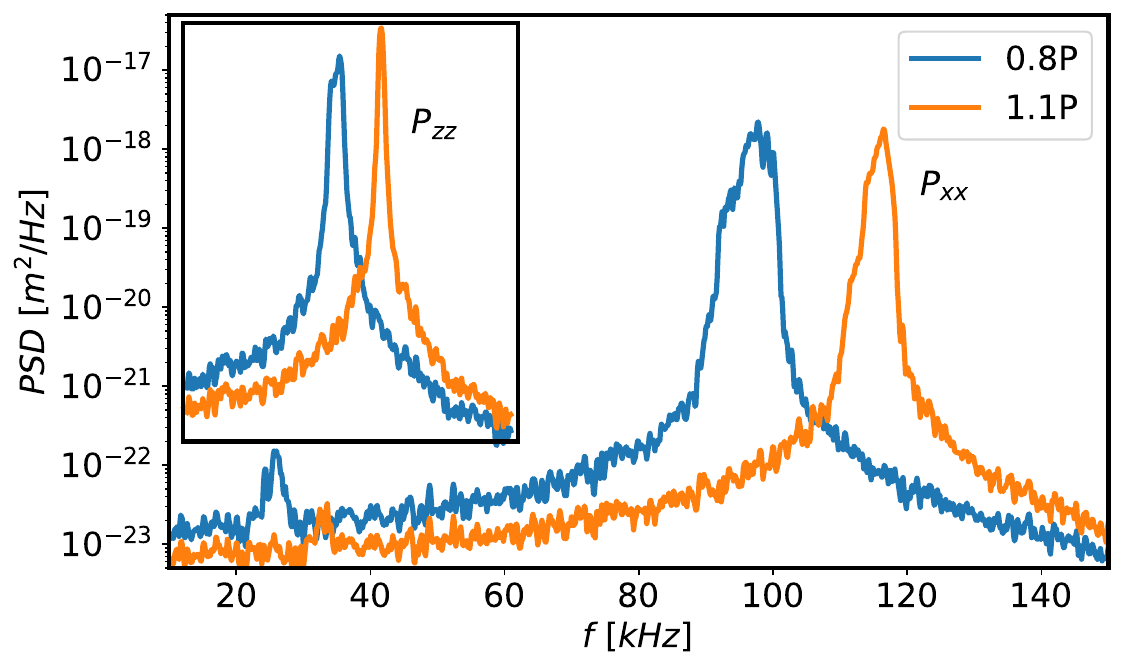}
    \caption{Simulated power spectral densities for a levitated particle with nominal radius $R=\SI{78}{\nano\meter}$ without feedback cooling. The observed mechanical frequencies $\Omega_z = 2\pi \times \SI{38}{\kilo\hertz}$,  $\Omega_{x,y} = 2\pi \times \SI{104}{\kilo\hertz}$ and sideband peaks at $\SI{24.2}{\kilo\hertz}$ are in good agreement with prediction by numerical calculations.}
    \label{fig:simulated_PSD}
\end{figure}

Figure \ref{fig:simulated_PSD} shows the simulated power spectral density (PSD) in the $x$ and $z$ (inset) directions over the same plotting interval at a pressure of $4 \times 10^{-2}$ mbar.
The simulated PSD is calculated using a stochastic Runge-Kutta method, taken over $500$ ms and rescaled to match the sample frequency of the experiment. Feedback cooling forces are not included.

The optical forces are calculated through the beam intensity generated by Eq. \ref{eq:E_field} for $\beta_0=0.4$ by changing the nominal power $P$ used in the experiment to more clowely match our experimental results. 
The blue curve is the PSD calculated for $0.8 P$ and orange is for $1.1 P$. These plots show how the trap frequencies vary with the change in beam power and the emergence of the nonlinear peaks. 

The main peaks are in good agreement with experimental values $\Omega_{x,y}$ and $\Omega_z$. %Noise floor is up to two orders of magnitude lower than the that found in our experiment which comes from the fact...... experiment has electronic noise and pxx and pzz
The sideband peaks observed in Fig. \ref{fig:PSD} are also present at $2\pi \times \SI{ 26.0}{\kilo \hertz}$ (blue) and $2\pi \times \SI{33.6}{\kilo \hertz}$ (orange). For the lower power calculated, we observe that the transverse peaks become wider and less resolved around the peak, which together with stronger sideband signals, indicates our experimental data lie within a parameter range with strong nonlinearities.

\newpage
\bibliography{main.bib}

\end{document}